\documentclass{iopart}
\usepackage{iopams}
\newcommand{\uD}{\mathrm{D}}

\begin{document}

 \title[Eddy~diffusivity of~short-correlated~random~flows]{Combined~role of~molecular~diffusion, mean~streaming and~helicity in~the~eddy~diffusivity of~short-correlated~random~flows}
 \author{Marco Martins Afonso$^1$, Andrea Mazzino$^2$ and S\'{\i}lvio Gama$^1$}
 \address{$^1$ Centro de Matem\'atica da Universidade do Porto, Rua do Campo Alegre 687, 4169-007 Porto, Portugal}
 \address{$^2$ DICCA, University of Genova, Via Montallegro 1, 16145 Genova, Italy and INFN \& CINFAI, Genova Section, Via Dodecaneso 33, 16146 Genova, Italy}
 \ead{marcomartinsafonso@hotmail.it}
 
 \begin{abstract}
  We analytically investigate the effective-diffusivity tensor of a tracer particle in a fluid flow endowed with a short correlation time.
  By means of functional calculus and a multiscale expansion, we write down the main contributions to the eddy diffusivity due to each single
  physical effect and to their interplays. Namely, besides molecular diffusivity and a constant uniform mean streaming, we take into account
  the possibility for the (incompressible, Gaussian, stationary, homogeneous, isotropic) turbulent fluctuations to break parity invariance.
  With respect to the classical turbulence-driven diffusivity amplification for delta-correlated flows, we find that the presence of a short
  temporal correlation induces a diminution even when coupled with such effects, with two principal exceptions. Notably, the diffusivity is ---
  perturbatively --- enlarged not only by the helical contribution itself, but also by the interference between molecular diffusion and mean flow.
 \end{abstract}
 
 
 \section{Introduction}
 
 Turbulent transport is a problem of remarkable importance in many situations, ranging from practical applications to pure sciences,
 where important achievements became possible thanks to the interplay between statistical-mechanics methods and classical fluid-dynamics analysis
 \cite{MK99,ALM00,AHMM00,W00,AAMMR01,FGV01,CM98,CMMV99}.
 A relevant limit (in short, the asymptotic limit) in turbulent transport is the one characterized by the double limit of very large spatial/temporal
 scales of the observed (transport) phenomena, with respect to the typical scales characterizing the advecting velocity field. By means of standard
 central-limit arguments, it is possible to see that the large-scale transport can be described in terms of an effective-diffusion equation,
 where the molecular (bare) diffusivity is replaced by a renormalized tensor named eddy diffusivity \cite{F87,F95}.
 This latter can be calculated (at least at a formal level) in terms of a formal multiple-scale expansion \cite{F87,BCVV95,M97}.
 Although, in general, exact explicit expressions for the eddy diffusivity are not available, approximate expressions have been proposed
 \cite{MMV05,CMMV06,BMAM15}.

 The present paper deals with turbulent transport in the asymptotic limit where the advecting velocity field is quasi delta-correlated in time.
 In the perfectly delta-correlated case it is well known that the eddy diffusivity can be computed exactly, and its expression is independent of
 the mean streaming, of the molecular diffusion, and of possible contributions arising from helicity. Our aim here is to investigate such a dependence
 once a small correlation time is allowed in the advecting velocity field. The presence of a small correlation time naturally leads to the definition
 of a small expansion coefficient, in terms of which a consistent perturbative approach can be carried out.

 The idea of our approach is to combine functional-calculus techniques with an exact result from the multiple-scale expansion obtained by
 \cite{F87,BCVV95} to compute the eddy diffusivity, with the final aim of arriving at exact (even if perturbative) expressions for the eddy diffusivity,
 through which the interplay of a mean streaming in the carrier flow, helicity in the turbulent fluctuations and molecular diffusion can be isolated.

 The paper is organized as follows. In section \ref{equ} we sketch the problem under investigation, we specify our assumptions on the flow
 and we enounce the entire final result. In section \ref{int} we discuss the role of each physical effect into play and how they interfere
 to build up the eddy diffusivity. Conclusions and perspectives follow in section \ref{con}. The appendix \ref{app} is devoted to show the
 details of the calculation and to recall the mathematical tools employed.

 \section{Equations} \label{equ}
 
 The auxiliary equation (or ``cell problem'' \cite{PS05,PS07}), in the presence of a constant uniform mean flow $\bi{U}$ \cite{HM94,FN10}
 and of molecular diffusivity $\kappa$, reads \cite{M97,MMV05,CMMV06,BML16}:
 \begin{equation} \label{ae}
  \partial_t\bi{w}(\bi{x},t)+[\bi{U}+\bi{v}(\bi{x},t)]\cdot\bpartial\bi{w}(\bi{x},t)-\kappa\partial^2\bi{w}(\bi{x},t)=-\bi{v}(\bi{x},t)\;.
 \end{equation}
 The random velocity fluctuations \cite{AM91,CFPV91} are assumed as incompressible,
 \begin{equation} \label{if}
  \bpartial\cdot\bi{v}(\bi{x},t)=0\;,
 \end{equation}
 and Gaussian, with zero mean,
 \begin{equation} \label{zm}
  \langle\bi{v}(\bi{x},t)\rangle=\mathbf{0}\;,
 \end{equation}
 and stationary--homogeneous--isotropic pair correlations:
 \begin{equation} \label{pc}
  \langle v_{\alpha}(\bi{x},t)v_{\beta}(\bi{x}',t')\rangle=\mathcal{V}_{\alpha\beta}(\bi{x}'-\bi{x})V(t'-t)\;.
 \end{equation}
 The exact form of the temporal part $V(\Delta t)$ is unessential as long as the correlation time $T$ is short,
 namely much smaller than the smallest characteristic advection time $\mathcal{L}/\mathcal{U}$
 (built from typical length and speed scales, $\mathcal{L}$ and $\mathcal{U}$, to be introduced shortly);
 to perform explicit calculations we will implement two specific forms, a decaying exponential and the rectangular function,
 which are both representatives of the Dirac delta in the limit of vanishing $T$ and are always non-negative.
 For what concerns the spatial part $\mathcal{V}_{\alpha\beta}(\bi{r})$ \cite{MY75},
 we can e.g.\ keep in mind an expression based on a Gaussian of width $L$ with prefactor $D_0$ (proportional to the turbulent kinetic energy),
 possibly modulated by a cosinusoid of wave number $k$ to take into account the presence of recirculation areas
 (the case $k=0$ being the ``classical'' situation without such zones).
 The mean flow $\bi{U}$ being uniform, the above-mentioned characteristic length scale $\mathcal{L}$ must be deduced
 from the turbulent fluctuations, and for our purpose it corresponds to the smaller between $L$ and the wave length $\ell\equiv2\pi/k$.
 About the aforementioned typical speed scale, it can derive from either the average streaming or the random perturbation.
 As no assumption is in principle made on the relative magnitude of the fields $\bi{U}$ and $\bi{v}$ \cite{M97,MMV05,CMMV06},
 to build the smallest advective time one must choose $\mathcal{U}$ as the maximum between the modulus $U$ and the root-mean-square value
 $\sqrt{D_0/T}$. Consequently, our condition on $T$ reads:
 \begin{equation} \label{ct}
  \frac{\min(L,\ell)}{U}\gg T\ll\frac{\min(L^2,\ell^2)}{D_0}\;.
 \end{equation}
 
 A formal temporal integration of (\ref{ae}) gives:%
 \footnote{All temporal integrals without explicit integration bounds are on the whole real axis,
  but of course the presence of a Heaviside theta introduces a (in this case, upper) bound.}
 \begin{eqnarray} \label{fi}
  \bi{w}(\bi{x},t)-\bi{w}(\bi{x},-\infty)=\\
  =-\!\int_{-\infty}^t\rmd\tau\,\{[\bi{U}+\bi{v}(\bi{x},\tau)]\cdot\bpartial\bi{w}(\bi{x},\tau)-\kappa\partial^2\bi{w}(\bi{x},\tau)+\bi{v}(\bi{x},\tau)\}\nonumber\\
  =\!\int\!\rmd\tau\,\theta(t-\tau)\{-\bi{v}(\bi{x},\tau)\cdot[\bpartial\bi{w}(\bi{x},\tau)+\mathtt{I}]-(\bi{U}\cdot\bpartial-\kappa\partial^2)\bi{w}(\bi{x},\tau)\}\nonumber\;.
 \end{eqnarray}
 From (\ref{ae}), notice that the auxiliary field can be assumed as having zero average,
 because this latter obeys an unforced (due to (\ref{zm})) advection--diffusion equation:
 \begin{equation} \label{za}
  \langle\bi{w}(\bi{x},t)\rangle=\mathbf{0}\;.
 \end{equation}

 The expression for the eddy-diffusivity tensor of tracer particles is \cite{BCVV95,AV95,M97,MK99}:
 \begin{equation} \label{ed}
  \mathcal{K}_{\alpha\beta}=\kappa\delta_{\alpha\beta}-\frac{\langle v_{\alpha}w_{\beta}+v_{\beta}w_{\alpha}\rangle}{2}\;.
 \end{equation}
 Let us then compute the quantity
 \begin{equation} \label{vw}\fl
  \langle v_{\alpha}(\bi{x},t)w_{\beta}(\bi{x},t)\rangle=\!\int\!\rmd\bi{x}^*\!\int\!\rmd t^*\,\langle v_{\alpha}(\bi{x},t)v_{\sigma}(\bi{x}^*,t^*)\rangle\left\langle\frac{\uD w_{\beta}(\bi{x},t)}{\uD v_{\sigma}(\bi{x}^*,t^*)}\right\rangle\;,
 \end{equation}
 which --- postponing the details to the appendix --- results into:
 \begin{eqnarray} \label{fr}\fl
  -\langle v_{\alpha}(\bi{x},t)w_{\beta}(\bi{x},t)\rangle\nonumber\\\fl
  =\!\int\!\rmd t^*\,\theta(t-t^*)\langle v_{\alpha}(\bi{x},t)v_{\beta}(\bi{x},t^*)\rangle\\\fl
  \quad+\!\int\!\rmd t'\!\int\!\rmd t''\!\int\!\rmd t^*\,\theta(t-t')\theta(t'-t'')\theta(t''-t^*)\langle v_{\alpha}(\bi{x},t)\partial_{\mu}\partial_{\nu}v_{\beta}(\bi{x},t^*)\rangle\langle v_{\mu}(\bi{x},t')v_{\nu}(\bi{x},t'')\rangle+\ldots\nonumber\\\fl
  \quad+\kappa\left(\!\int\!\rmd t'\!\int\!\rmd t^*\,\theta(t-t')\theta(t'-t^*)\langle v_{\alpha}(\bi{x},t)\partial^2v_{\beta}(\bi{x},t^*)\rangle+\ldots\right)\nonumber\\\fl
  \quad+U_{\mu}U_{\nu}\left(\!\int\!\rmd t'\!\int\!\rmd t''\!\int\!\rmd t^*\,\theta(t-t')\theta(t'-t'')\theta(t''-t^*)\langle v_{\alpha}(\bi{x},t)\partial_{\mu}\partial_{\nu}v_{\beta}(\bi{x},t^*)\rangle+\ldots\right)\nonumber\\\fl
  \quad+3\kappa U_{\mu}U_{\nu}\left(\!\int\!\rmd t'\!\int\!\rmd t''\!\int\!\rmd t'''\!\int\!\rmd t^*\,\theta(t-t')\theta(t'-t'')\theta(t''-t''')\theta(t'''-t^*)\langle v_{\alpha}(\bi{x},t)\partial_{\mu}\partial_{\nu}\partial^2v_{\beta}(\bi{x},t^*)\rangle+\ldots\right).\nonumber
 \end{eqnarray}
 By rearranging the temporal integrals and symmetrizing on $\alpha\leftrightarrow\beta$ (which is a trivial operation), one obtains:
 \begin{eqnarray} \label{rs}\fl
  \mathcal{K}_{\alpha\beta}=&\kappa\delta_{\alpha\beta}+\mathcal{V}_{\alpha\beta}(\mathbf{0})\!\int_0^{+\infty}\rmd\tau\,V(\tau)+\mathcal{V}_{\mu\nu}(\mathbf{0})\partial_{\mu}\partial_{\nu}\mathcal{V}_{\alpha\beta}|_{\bi{r}=\mathbf{0}}\!\int_0^{+\infty}\rmd\tau\,V(\tau)\!\int_0^{\tau}\rmd\tau'\,(\tau-\tau')V(\tau')\nonumber\\\fl
  &+\kappa\partial^2\mathcal{V}_{\alpha\beta}|_{\bi{r}=\mathbf{0}}\!\int_0^{+\infty}\rmd\tau\,\tau V(\tau)+\frac{1}{2}U_{\mu}U_{\nu}\partial_{\mu}\partial_{\nu}\mathcal{V}_{\alpha\beta}|_{\bi{r}=\mathbf{0}}\!\int_0^{+\infty}\!\rmd\tau\,\tau^2V(\tau)\nonumber\\\fl
  &+\frac{1}{2}\kappa U_{\mu}U_{\nu}\partial_{\mu}\partial_{\nu}\partial^2\mathcal{V}_{\alpha\beta}|_{\bi{r}=\mathbf{0}}\!\int_0^{+\infty}\rmd\tau\,\tau^3V(\tau)+\textrm{h.o.t.}\;.
 \end{eqnarray}

 Notice that expression (\ref{rs}) only holds if the velocity field is parity invariant,
 i.e.\ if the spatial part of its pair correlation in (\ref{pc}) can be expressed as:
 \begin{equation} \label{pi}
  \mathcal{V}_{\alpha\beta}(\bi{r})=A(r)\delta_{\alpha\beta}+B(r)\frac{r_{\alpha}r_{\beta}}{r^2}\;.
 \end{equation}
 As a consequence, odd-order derivatives of $\mathcal{V}_{\alpha\beta}(\bi{r})$ vanish when computed in $\bi{r}=\mathbf{0}$,
 while for the (spatial part of) relevant even-order counterparts one finds:
 \begin{equation*}
  \langle v_{\alpha}v_{\beta}\rangle=\delta_{\alpha\beta}A(0)\;,
 \end{equation*}
 \begin{equation*}
  \langle v_{\alpha}\partial_{\mu}\partial_{\nu}v_{\beta}\rangle=\delta_{\alpha\beta}\delta_{\mu\nu}\left.\frac{A'(r)}{r}\right|_{r=0}+(\delta_{\alpha\mu}\delta_{\beta\nu}+\delta_{\alpha\nu}\delta_{\beta\mu})\left.\frac{B(r)}{r^2}\right|_{r=0}\;,
 \end{equation*}
 \begin{eqnarray*}
  \langle v_{\alpha}\partial_{\mu}\partial_{\nu}\partial^2v_{\beta}\rangle=&(d+2)\delta_{\alpha\beta}\delta_{\mu\nu}\left[\frac{A''(r)}{r^2}-\frac{A'(r)}{r^3}\right]_{r=0}+[2\delta_{\alpha\beta}\delta_{\mu\nu}\qquad\\
  &+(d+4)(\delta_{\alpha\mu}\delta_{\beta\nu}+\delta_{\alpha\nu}\delta_{\beta\mu})]\left[\frac{B'(r)}{r^3}-2\frac{B(r)}{r^4}\right]_{r=0}\;.
 \end{eqnarray*}
 
 \subsection{Parity-breaking flows}
 
 It was shown in \cite{BCVV95} that different contributions due to helicity (the averaged scalar product between velocity and its curl, i.e.\ vorticity)
 arise if the flow is invariant under translations and rotations but not under reflections, viz.\ if also an addend showing the fully-antisymmetric
 pseudotensor $\varepsilon_{\alpha\beta\gamma}$ is allowed on the right-hand side of (\ref{pi}) in the 3D case, so that \cite{L08}:
 \begin{equation} \label{pb}
  \mathcal{V}_{\alpha\beta}(\bi{r})=A(r)\delta_{\alpha\beta}+B(r)\frac{r_{\alpha}r_{\beta}}{r^2}+C(r)\varepsilon_{\alpha\beta\gamma}\frac{r_{\gamma}}{r}\;.
 \end{equation}
 The appearance of the new terms is due to the fact that, while even-order derivatives of $\mathcal{V}_{\alpha\beta}(\bi{r})$ computed at
 $\bi{r}=\mathbf{0}$ keep unchanged, now also their odd-order counterparts do not vanish. Namely, from (\ref{pb}) for the spatial part one has:
 \begin{equation*}
  \langle v_{\alpha}\partial_{\lambda}v_{\beta}\rangle=\varepsilon_{\alpha\beta\lambda}\left.\frac{C(r)}{r}\right|_{r=0}\;,
 \end{equation*}
 \begin{equation*}
  \langle v_{\alpha}\partial_{\lambda}\partial_{\mu}\partial_{\nu}v_{\beta}\rangle=(\varepsilon_{\alpha\beta\lambda}\delta_{\mu\nu}+\varepsilon_{\alpha\beta\mu}\delta_{\lambda\nu}+\varepsilon_{\alpha\beta\nu}\delta_{\lambda\mu})\left[\frac{C'(r)}{r^2}-\frac{C(r)}{r^3}\right]_{r=0}\!.
 \end{equation*}
 Notice that $A(0)>0$ (a measure of the turbulent kinetic energy), while $B(0)=0=C(0)$.

 Given first of all that the $O(1)$ delta-correlated-like result (first line of (\ref{fr})) cannot be altered by this modification,
 at leading-correction order two new contributions appear, but the only relevant is
 \begin{equation} \label{hf}\fl
  +\!\int\!\rmd t'\!\int\!\rmd t''\!\int\!\rmd t^*\,\theta(t-t')\theta(t'-t'')\theta(t''-t^*)\langle v_{\alpha}(\bi{x},t)\partial_{\mu}v_{\nu}(\bi{x},t^*)\rangle\langle v_{\mu}(\bi{x},t')\partial_{\nu}v_{\beta}(\bi{x},t'')\rangle
 \end{equation}
 (which can be recast as $\displaystyle\partial_{\mu}\mathcal{V}_{\alpha\nu}|_{\bi{r}=\mathbf{0}}\partial_{\nu}\mathcal{V}_{\mu\beta}|_{\bi{r}=\mathbf{0}}\!\int_0^{+\infty}\rmd\tau\,V(\tau)\!\int_0^{\tau}\rmd\tau'\,(\tau-\tau')V(\tau')$).
 Expression (\ref{hf}) can be considered as modifying the second line of the right-hand side of
 (\ref{fr}), but with a different distribution of the spatial derivatives which is finite only for flows not symmetric under reflections \cite{BCVV95}.\\
 The other main contribution, appearing in (\ref{fr}) but \emph{not} in (\ref{rs}), is
 \begin{equation} \label{mf}\fl
  -U_{\lambda}\!\int\!\rmd t'\!\int\!\rmd t^*\,\theta(t-t')\theta(t'-t^*)\langle v_{\alpha}(\bi{x},t)\partial_{\lambda}v_{\beta}(\bi{x},t^*)\rangle=-U_{\lambda}\partial_{\lambda}\mathcal{V}_{\alpha\beta}|_{\bi{r}=\mathbf{0}}\!\int_0^{+\infty}\!\rmd\tau\,\tau V(\tau)\;.
 \end{equation}
 Actually, (\ref{mf}) does not induce any correction to the eddy diffusivity because it is proportional to $\varepsilon_{\alpha\beta\lambda}$ and
 antisymmetric in the indices $\alpha\leftrightarrow\beta$, so that a cancellation takes place when symmetrizing according to (\ref{ed}).\\
 Analyzing the following order, one obtains the leading coupling between molecular diffusivity and helicity
 popping up on the right-hand side of (\ref{fr}),
 \begin{eqnarray} \label{dh}\fl
  +\kappa\!\int\!\rmd t'\!\int\!\rmd t''\!\int\!\rmd t'''\!\int\!\rmd t^*\,\theta(t-t')\theta(t'-t'')\theta(t''-t''')\theta(t'''-t^*)\times\\\fl
  \quad\times(\langle v_{\alpha}(\bi{x},t)\partial_{\lambda}\partial^2v_{\gamma}(\bi{x},t''')\rangle\langle v_{\lambda}(\bi{x},t')\partial_{\gamma}v_{\beta}(\bi{x},t^*)\rangle+\langle v_{\alpha}(\bi{x},t)\partial_{\lambda}v_{\gamma}(\bi{x},t''')\rangle\langle v_{\lambda}(\bi{x},t')\partial_{\gamma}\partial^2v_{\beta}(\bi{x},t^*)\rangle\nonumber\\\fl
  \qquad+\langle v_{\alpha}(\bi{x},t)\partial_{\lambda}\partial^2v_{\gamma}(\bi{x},t''')\rangle\langle v_{\lambda}(\bi{x},t'')\partial_{\gamma}v_{\beta}(\bi{x},t^*)\rangle+\langle v_{\alpha}(\bi{x},t)\partial_{\lambda}v_{\gamma}(\bi{x},t'')\rangle\langle v_{\lambda}(\bi{x},t')\partial_{\gamma}\partial^2v_{\beta}(\bi{x},t^*)\rangle)\nonumber
 \end{eqnarray}
 (i.e.
 \begin{equation*}\fl
  \kappa\left(\partial_{\lambda}\partial^2\mathcal{V}_{\alpha\gamma}\times\partial_{\gamma}\mathcal{V}_{\lambda\beta}+\partial_{\lambda}\mathcal{V}_{\alpha\gamma}\times\partial_{\gamma}\partial^2\mathcal{V}_{\lambda\beta}\right)_{\bi{r}=\mathbf{0}}\!\int_0^{+\infty}\rmd\tau\left\{\!\int_0^{\tau}\rmd\tau'\,\tau\tau'+\!\int_{\tau}^{+\infty}\rmd\tau'\,\tau^2\right\}V(\tau)V(\tau')
 \end{equation*}
 equivalently),
 but --- as provable by means of lengthy algebra --- not yet the leading term of the coupling between mean flow and helicity.\\
 As it is nevertheless interesting to write down this latter, we have pushed the calculation at the next-to-following order and,
 among all possible terms, we can state that only few of them survive and represent this coupling.
 Namely, on the right-hand side of (\ref{fr}) there appears
 \begin{eqnarray} \label{mh}\fl
  +U_{\mu}U_{\nu}\!\int\!\rmd t'\!\int\!\rmd t''\!\int\!\rmd t'''\!\int\!\rmd t^{\dag}\!\int\!\rmd t^*\,\theta(t-t')\theta(t'-t'')\theta(t''-t''')\theta(t'''-t^{\dag})\theta(t^{\dag}-t^*)\times\\\fl
  \quad\times(\langle v_{\alpha}(\bi{x},t)\partial_{\mu}\partial_{\nu}\partial_{\lambda}v_{\gamma}(\bi{x},t^{\dag})\rangle\langle v_{\lambda}(\bi{x},t')\partial_{\gamma}v_{\beta}(\bi{x},t^*)\rangle+\langle v_{\alpha}(\bi{x},t)\partial_{\lambda}v_{\gamma}(\bi{x},t^{\dag})\rangle\langle v_{\lambda}(\bi{x},t')\partial_{\mu}\partial_{\nu}\partial_{\gamma}v_{\beta}(\bi{x},t^*)\rangle\nonumber\\\fl
  \qquad+\langle v_{\alpha}(\bi{x},t)\partial_{\mu}\partial_{\nu}\partial_{\lambda}v_{\gamma}(\bi{x},t^{\dag})\rangle\langle v_{\lambda}(\bi{x},t'')\partial_{\gamma}v_{\beta}(\bi{x},t^*)\rangle+\langle v_{\alpha}(\bi{x},t)\partial_{\lambda}v_{\gamma}(\bi{x},t''')\rangle\langle v_{\lambda}(\bi{x},t')\partial_{\mu}\partial_{\nu}\partial_{\gamma}v_{\beta}(\bi{x},t^*)\rangle\nonumber\\\fl
  \qquad+\langle v_{\alpha}(\bi{x},t)\partial_{\mu}\partial_{\nu}\partial_{\lambda}v_{\gamma}(\bi{x},t^{\dag})\rangle\langle v_{\lambda}(\bi{x},t''')\partial_{\gamma}v_{\beta}(\bi{x},t^*)\rangle+\langle v_{\alpha}(\bi{x},t)\partial_{\lambda}v_{\gamma}(\bi{x},t'')\rangle\langle v_{\lambda}(\bi{x},t')\partial_{\mu}\partial_{\nu}\partial_{\gamma}v_{\beta}(\bi{x},t^*)\rangle)\nonumber
 \end{eqnarray}
 (or
 \begin{equation*}\fl
  \frac{1}{2}U_{\mu}U_{\nu}\left(\partial_{\mu}\partial_{\nu}\partial_{\lambda}\mathcal{V}_{\alpha\gamma}\times\partial_{\gamma}\mathcal{V}_{\lambda\beta}+\partial_{\lambda}\mathcal{V}_{\alpha\gamma}\times\partial_{\mu}\partial_{\nu}\partial_{\gamma}\mathcal{V}_{\lambda\beta}\right)_{\bi{r}=\mathbf{0}}\!\int_0^{+\infty}\rmd\tau\left\{\!\int_0^{\tau}\rmd\tau'\,\tau'\tau^2+\!\int_{\tau}^{+\infty}\rmd\tau'\,\tau^3\right\}V(\tau)V(\tau')
 \end{equation*}
 equivalently).
 Expressions (\ref{dh}) and (\ref{mh}) can be shown to be already symmetric in their two free indices,
 and thus appear in the same form also in the eddy diffusivity (\ref{rs}).\\
 For the sake of completeness, even if we do not report it here, we point out that the leading coupling between
 molecular diffusivity -- mean flow -- helicity is captured by further increasing the order of investigation by one.
 In summary, leading corrections due to helical terms
 take place at an order higher by unity with respect to the corresponding non-helical contributions.

 \section{Interference} \label{int}
 
 Our main objective is to show the leading contribution due to each physical effect into play,
 and of course the interplay between these latter. To make our analysis more precise and quantitative,
 it is desirable to express the results in terms of \emph{interference} \cite{MV97,CCMVV98,MC99,MAMM12}.
 Let us focus first on the helicity-free case, i.e.\ on reflection-symmetric flows.
 For the sake of notational simplicity, the eddy diffusivity computed in (\ref{rs}) (before introducing
 the possibility of parity breakup) is denoted as $K_{\alpha\beta}\equiv\mathcal{K_{\alpha\beta}}|_{C(r)=0}$.
 
 To quantify our results, let us specify explicit forms for the turbulent-velocity correlations.
 For the temporal part, the two simplest representatives of the Dirac delta are
 \begin{equation} \label{sc}
  V(\Delta t)=\ (\textrm{I})\;\frac{\mathrm{e}^{-|\Delta t|/T}}{2T},\ (\textrm{II})\;\frac{\theta(T-|\Delta t|)}{2T}
 \end{equation}
 (note that this choice is convenient and possible because, in the time domain, we have to perform only integrals but no derivative,
 so the discontinuity at the step and the non-differentiability of the modulus do not rise any problem).
 About the space domain, we notice that according to our formulas spatial derivatives have to be performed at zero separation.
 Consequently, a suitable choice for what is usually dubbed ``longitudinal structure function'' \cite{MY75} is
 \begin{equation} \label{mg}
  D(r)=D_0\mathrm{e}^{-r^2/2L^2}\cos(kr)\;.
 \end{equation}
 The incompressibility constraint allows one to simply deduce the ``normal'' counterpart and then the full tensor
 $\mathcal{V}_{\alpha\beta}(\bi{r})$ in (\ref{pi}), with
 \begin{equation}\fl
  A(r)=D(r)+\frac{r}{d-1}D'(r)=\frac{D_0}{d-1}\mathrm{e}^{-r^2/2L^2}\left[\left(d-1-\frac{r^2}{L^2}\right)\cos(kr)-kr\sin(kr)\right]
 \end{equation}
 and 
 \begin{equation}\fl
  B(r)=-\frac{r}{d-1}D'(r)=\frac{D_0}{d-1}\mathrm{e}^{-r^2/2L^2}\left[\frac{r^2}{L^2}\cos(kr)+kr\sin(kr)\right]\;.
 \end{equation}
 
 \paragraph{Turbulent contribution} The first obvious result is about the role of the turbulent flow. If all other effects (molecular diffusion
 and mean flow) are switched off, the eddy diffusivity (\ref{rs}) keeps a diagonal form:
 \begin{equation} \label{td}\fl
  K_{\alpha\beta}|_{\kappa=U=0}=\delta_{\alpha\beta}A(0)\!\int_0^{+\infty}\rmd\tau\,V(\tau)\left\{1+\left[d\frac{A'}{r}+2\frac{B}{r^2}\right]_{r=0}\!\int_0^{\tau}\rmd\tau'\,(\tau-\tau')V(\tau')\right\}+\textrm{h.o.t.}\;.
 \end{equation}
 An incompressible turbulent flow, as is well known, enhances the eddy diffusivity. However, it can easily be shown that this enhancement
 is maximum for delta-correlated flows (where only the former term in curly braces survives), because in the presence of time correlation
 the latter term --- negative because of the factor in square brackets --- contributes to an attenuation of this effect \cite{BCVV95}.
 Indeed, for any admissible form of the temporal correlation, it is evident that the former term in (\ref{td}) is $O(T^0)$ and positive,
 and exactly gives rise to the equivalent of Taylor's formula for delta-correlated flows:
 \begin{equation} \label{dc}
  \mathcal{K}_{\alpha\beta}|_{T=0}=\kappa\delta_{\alpha\beta}+\frac{1}{2}\mathcal{V}_{\alpha\beta}(\mathbf{0})=\delta_{\alpha\beta}\left(\kappa+\frac{D_0}{2}\right)\;.
 \end{equation}
 (Notice that, due to the chaoticity of the random flow under consideration, the limit case $\kappa=0$ can be analyzed without any singularity.)
 On the contrary, the latter term in (\ref{td}) evaluates explicitly through (\ref{sc}I) and (\ref{mg}) as
 \begin{equation*} 
  -\frac{1}{8}D_0^2(d+2)\left(k^2+\frac{1}{L^2}\right)\delta_{\alpha\beta}T\;,
 \end{equation*}
 with the numerical prefactor becoming $-1/24$ for (\ref{sc}II).
 It contributes \emph{negatively} at $O(T^1)$ to the eddy diffusivity, is larger in magnitude in 3D than in 2D,
 and becomes relevant when either $k^2$ or $L^{-2}$ is non-negligible (but anyhow small,
 always in a perturbative spirit: see (\ref{ct})) with respect to $(D_0T)^{-1}$.
 We point out that this contribution, as well as the ones shown in the next paragraphs, can get the opposite sign if the assumption of non-negativity
 for $V(\Delta t)$ is released \cite{M97} --- e.g.\ by considering a sinc function or a modulated decaying exponential.
 
 \paragraph{Role of molecular diffusivity} Secondly, let us study the role of molecular diffusivity in the absence of mean flow.
 If, from the full non-helical expression of the eddy diffusivity computed at $U=0$, we subtract both the expression
 in the absence of molecular diffusion and the one in the absence of turbulent flow (which trivially reduces to $\kappa\delta_{\alpha\beta}$), we get:
 \begin{equation*}\fl
  K_{\alpha\beta}|_{U=0}-K_{\alpha\beta}|_{\kappa=U=0}-K_{\alpha\beta}|_{D(r)=U=0}=\kappa\delta_{\alpha\beta}\left[d\frac{A'}{r}+2\frac{B}{r^2}\right]_{r=0}\!\int_0^{+\infty}\rmd\tau\,\tau V(\tau)+\textrm{h.o.t.}\;.
 \end{equation*}
 Again because of the negativity of the factor in square brackets, we can assert that --- at leading order $O(T^1)$ ---
 molecular diffusion and turbulence interfere \emph{destructively} in building up the eddy diffusivity. The exact result is
 \begin{equation*}
  -\frac{1}{2}\kappa D_0(d+2)\left(k^2+\frac{1}{L^2}\right)\delta_{\alpha\beta}T\;,
 \end{equation*}
 for (\ref{sc}I) and (\ref{mg}), with the numerical prefactor becoming $-1/4$ for (\ref{sc}II).
 
 \paragraph{Role of mean flow} As a third point, we look for the effect of the mean flow in the absence of molecular diffusivity:
 \begin{equation*}\fl
  K_{\alpha\beta}|_{\kappa=0}-K_{\alpha\beta}|_{\kappa=U=0}=\left(\frac{1}{2}U^2\delta_{\alpha\beta}\frac{A'}{r}+U_{\alpha}U_{\beta}\frac{B}{r^2}\right)_{r=0}\!\int_0^{+\infty}\!\rmd\tau\,\tau^2V(\tau)+\textrm{h.o.t.}\;.
 \end{equation*}
 Here, the matrix in round parentheses shows positive off-diagonal terms which however can be eliminated by moving to the principal axes,
 but its trace is negative. The quantitative result for (\ref{sc}I) and (\ref{mg}) is
 \begin{equation*}
  -\frac{1}{2}\frac{D_0}{d-1}[(d+1)U^2\delta_{\alpha\beta}-2U_{\alpha}U_{\beta}]\left(k^2+\frac{1}{L^2}\right)T^2\;,
 \end{equation*}
 with the numerical prefactor becoming $-1/12$ for (\ref{sc}II).
 Therefore, the conclusion about the leading role of the mean flow points in the same direction
 of \emph{depletion} of the eddy diffusivity, but with three remarks:
 1) the effect is here $O(T^2)$, and becomes perturbatively relevant when $T$ is non-negligible with respect to $U^2/D_0$;
 2) the depletion in the direction aligned with the mean flow is reduced with respect to the orthogonal one(s);
 3) of course it does not make any sense to compute $\mathcal{K}_{\alpha\beta}$ in the presence of mean flow and absence of turbulent fluctuations,
 because a deterministic (constant) flow alone introduces an advective feature but no diffusion at all.
 
 \paragraph{Coupling $\kappa$-$\bi{U}$} Lastly, it is also interesting to investigate the coupling between molecular diffusivity and mean flow
 (clearly in concourse with a turbulent flow):
 \begin{eqnarray*}\fl
  K_{\alpha\beta}-K_{\alpha\beta}|_{\kappa=0}-K_{\alpha\beta}|_{U=0}+K_{\alpha\beta}|_{\kappa=U=0}=\\\fl
  =\kappa\left\{\frac{1}{2}(d+2)U^2\delta_{\alpha\beta}\left[\frac{A''}{r^2}-\frac{A'}{r^3}\right]+(U^2\delta_{\alpha\beta}+(d+4)U_{\alpha}U_{\beta})\left[\frac{B'}{r^3}-2\frac{B}{r^4}\right]\right\}_{r=0}\!\int_0^{+\infty}\rmd\tau\,\tau^3V(\tau)+\textrm{h.o.t.}\;.
 \end{eqnarray*}
 The former square bracket is positive, while the latter is negative. If one computes the trace,
 a positive overall result is found, meaning that molecular diffusion and mean flow --- with the essential aid of turbulence ---
 interfere \emph{constructively} to build up the eddy diffusivity. As for the previous point, the effect (in this case, with the opposite sign
 and at $O(T^3)$) is larger in the direction(s) perpendicular to the mean flow than for the parallel one, and the negative result for non-diagonal terms
 can be discarded by changing the reference frame. The effective result for (\ref{sc}I) and (\ref{mg}) is
 \begin{equation*}
  \frac{1}{2}\kappa D_0\frac{d+4}{d-1}[(d+1)U^2\delta_{\alpha\beta}-2U_{\alpha}U_{\beta})]\left(k^4+\frac{6k^2}{L^2}+\frac{3}{L^4}\right)T^3\;,
 \end{equation*}
 with the numerical prefactor becoming $1/48$ for (\ref{sc}II). 
 
 \subsection{Parity-breaking flows}
 
 Let us now take helicity into account, i.e.\ $C(r)\neq0$ in $d=3$. In particular, following (\ref{mg}), a suitable choice is
 \begin{equation}\fl
  C(r)=c\sqrt{-rD(r)D'(r)}=cD_0\mathrm{e}^{-r^2/2L^2}\frac{r}{L}\sqrt{\cos^2(kr)+\frac{kL^2}{r}\cos(kr)\sin(kr)}\;.
 \end{equation}
 Even if $C(r)$ has the same physical dimensions as $A(r)$ and $B(r)$, and the helicity and energy spectra obey an important inequality
 in the Fourier space \cite{BCVV95,L08,D11}, the dimensional prefactor is in general different from $D_0$ and has been written here as $cD_0$.
 The non-dimensional constant $c$ vanishes for non-helical flows and always appears as squared in physically-relevant results,
 in accordance with the fact that its sign is meaningless since dependent on the choice of a positive rotation convention.
 
 \paragraph{Turbulent contribution} Computing the turbulent contribution to the eddy diffusivity again, one has:
 \begin{equation} \label{hc}\fl
  \mathcal{K}_{\alpha\beta}|_{\kappa=U=0}=\delta_{\alpha\beta}\!\int_0^{+\infty}\rmd\tau\,V(\tau)\left\{A(0)+\left[A\left(3\frac{A'}{r}+2\frac{B}{r^2}\right)+2\frac{C^2}{r^2}\right]_{r=0}\!\int_0^{\tau}\rmd\tau'\,(\tau-\tau')V(\tau')\right\}+\textrm{h.o.t.}\;.
 \end{equation}
 The new helical contribution is positive, as already shown in \cite{BCVV95}. Helicity itself thus provides an \emph{augmentation} of the eddy diffusivity
 at $O(T^1)$, e.g.\ for (\ref{sc}I) and (\ref{mg}) the third term in (\ref{hc}) (corresponding to (\ref{hf}) and absent in (\ref{dc})) is
 \begin{equation*} 
  \frac{c^2}{4}D_0^2\left(k^2+\frac{1}{L^2}\right)\delta_{\alpha\beta}T\;,
 \end{equation*}
 with the numerical prefactor becoming $c^2/12$ for (\ref{sc}II).

 \paragraph{Role of molecular diffusivity} The computation of the coupling between molecular diffusion and helicity from (\ref{dh}) gives:
 \begin{eqnarray*}\fl
  10\kappa\delta_{\alpha\beta}\left[\frac{C}{r}\left(\frac{C'}{r^2}-\frac{C}{r^3}\right)\right]_{r=0}\!\int_{-\infty}^t\rmd t'\!\int_{-\infty}^{t'}\rmd t''\!\int_{-\infty}^{t''}\rmd t'''\!\int_{-\infty}^{t'''}\rmd t^*\\\fl
  \qquad\{V(t'''-t)[V(t^*-t')+V(t^*-t'')]+[V(t''-t)+V(t'''-t)]V(t^*-t')\}\;.
 \end{eqnarray*}
 More specifically, imposing (\ref{sc}I) and (\ref{mg}), one gets
 \begin{equation*}
  -\frac{5c^2}{4}\kappa D_0^2\delta_{\alpha\beta}\left(2k^4+\frac{6k^2}{L^2}+\frac{3}{L^4}\right)T^2\;,
 \end{equation*}
 while for (\ref{sc}II) the numerical prefactor is $-25c^2/72$. Therefore, the interplay between molecular diffusion and turbulent fluctuations
 brings about a \emph{reduction} of the effective diffusivity even when the helical component is taken into account, but this time at $O(T^2)$.
 
 \paragraph{Role of mean flow} What is more interesting is the coupling between mean flow and helicity.
 Computing all the contributions in (\ref{mh}) and summing up, we are left with:
 \begin{eqnarray*}\fl
  (4U^2\delta_{\alpha\beta}-2U_{\alpha}U_{\beta})\left[\frac{C}{r}\left(\frac{C'}{r^2}-\frac{C}{r^3}\right)\right]_{r=0}\!\int_{-\infty}^t\rmd t'\!\int_{-\infty}^{t'}\rmd t''\!\int_{-\infty}^{t''}\rmd t'''\!\int_{-\infty}^{t'''}\rmd t^{\dag}\!\int_{-\infty}^{t^{\dag}}\rmd t^*\{V(t^{\dag}-t)\times\\\fl
  \quad\times[V(t^*-t')+V(t^*-t'')+V(t^*-t''')]+[V(t^{\dag}-t)+V(t''-t)+V(t'''-t)]V(t^*-t')\}\;.
 \end{eqnarray*}
 For the explicit benchmark correlations (\ref{sc}I) and (\ref{mg}), this becomes
 \begin{equation*}
  -\frac{7c^2}{24}D_0^2(2U^2\delta_{\alpha\beta}-U_{\alpha}U_{\beta})\left(2k^4+\frac{6k^2}{L^2}+\frac{3}{L^4}\right)T^3\;,
 \end{equation*}
 with the numerical prefactor becoming $-c^2/40$ for (\ref{sc}II). The conclusion is that the presence of a mean streaming always \emph{diminishes}
 the eddy diffusivity even when coupled with helical turbulence, but this effect is $O(T^3)$, and is relevant from a perturbative viewpoint
 when either $k^2$ or $L^{-2}$ is small-but-not-negligible (see (\ref{ct})) with respect to $(UT)^{-2}$.

 \section{Conclusions and perspectives} \label{con}
 
 We have analyzed the eddy diffusivity of a tracer particle in incompressible random flows,
 namely in those endowed with a homogeneous isotropic Gaussian probability density function.
 By means of functional calculus and of the Furutsu--Novikov--Donsker theorem,
 we have expressed our result in a power series in the (small) velocity correlation time $T$.
 Taking into account the presence of molecular diffusivity $\kappa$ and of a constant uniform mean streaming $\bi{U}$,
 and the possibility for the turbulent fluctuations to break parity invariance in the form
 of a non-zero helicity, our focus was on the role of each single effect as well as on their
 interplay in building up the effective-diffusivity tensor.\\
 Besides the ``classical'' turbulent contribution which augments diffusion
 by potentially many orders of magnitude, and acts at $O(T^0)$ also in the case of temporal
 delta-correlation, we have explicitly obtained the physical-space expressions
 --- already investigated in \cite{BCVV95} in the Fourier space --- of the two $O(T^1)$ corrections
 purely due to turbulence. They act in an opposite sense with respect to each other,
 as the term related to kinetic energy (through a Laplacian, i.e.\ an even-order derivative)
 decreases the diffusivity, while the one possibly arising from helicity (involving a curl, i.e.\ an odd-order derivative)
 increases it. The sweeping effect of a mean flow is always to deplete the eddy diffusivity,
 both when interacting at $O(T^2)$ with usual parity-invariant flows as already known,
 and when coupled with a helical component at $O(T^3)$. The role of molecular diffusivity points in the same direction
 (at $O(T^1)$ and $O(T^2)$ respectively), however it is worth mentioning that the coupling between $\kappa$ and $\bi{U}$
 interestingly enough amplifies the effective diffusion, even if as a subleading perturbative correction.\\
 Our global result, gleaning the non-helical expression (\ref{rs}) together with the leading helical corrections
 from (\ref{hf}), (\ref{dh}) and (\ref{mh}), is:
 \begin{eqnarray} \label{gr}\fl
  \mathcal{K}_{\alpha\beta}=&\kappa\delta_{\alpha\beta}+\mathcal{V}_{\alpha\beta}(\mathbf{0})\!\int_0^{+\infty}\rmd\tau\,V(\tau)\nonumber\\\fl
  &+(\mathcal{V}_{\mu\nu}\partial_{\mu}\partial_{\nu}\mathcal{V}_{\alpha\beta}+\partial_{\mu}\mathcal{V}_{\alpha\nu}\times\partial_{\nu}\mathcal{V}_{\mu\beta})_{\bi{r}=\mathbf{0}}\!\int_0^{+\infty}\rmd\tau\,V(\tau)\!\int_0^{\tau}\rmd\tau'\,(\tau-\tau')V(\tau')\nonumber\\\fl
  &+\kappa\partial^2\mathcal{V}_{\alpha\beta}|_{\bi{r}=\mathbf{0}}\!\int_0^{+\infty}\rmd\tau\,\tau V(\tau)+\frac{1}{2}U_{\mu}U_{\nu}\partial_{\mu}\partial_{\nu}\mathcal{V}_{\alpha\beta}|_{\bi{r}=\mathbf{0}}\!\int_0^{+\infty}\!\rmd\tau\,\tau^2V(\tau)\nonumber\\\fl
  &+\kappa\left(\partial_{\lambda}\partial^2\mathcal{V}_{\alpha\gamma}\times\partial_{\gamma}\mathcal{V}_{\lambda\beta}+\partial_{\lambda}\mathcal{V}_{\alpha\gamma}\times\partial_{\gamma}\partial^2\mathcal{V}_{\lambda\beta}\right)_{\bi{r}=\mathbf{0}}\times\nonumber\\\fl
  &\qquad\times\!\int_0^{+\infty}\rmd\tau\!\int_0^{+\infty}\rmd\tau'\,\tau\min(\tau,\tau')V(\tau)V(\tau')\nonumber\\\fl
  &+\frac{1}{2}U_{\mu}U_{\nu}\left(\partial_{\mu}\partial_{\nu}\partial_{\lambda}\mathcal{V}_{\alpha\gamma}\times\partial_{\gamma}\mathcal{V}_{\lambda\beta}+\partial_{\lambda}\mathcal{V}_{\alpha\gamma}\times\partial_{\mu}\partial_{\nu}\partial_{\gamma}\mathcal{V}_{\lambda\beta}\right)_{\bi{r}=\mathbf{0}}\times\nonumber\\\fl
  &\qquad\times\!\int_0^{+\infty}\rmd\tau\!\int_0^{+\infty}\rmd\tau'\,\tau^2\min(\tau,\tau')V(\tau)V(\tau')\nonumber\\\fl
  &+\frac{1}{2}\kappa U_{\mu}U_{\nu}\partial_{\mu}\partial_{\nu}\partial^2\mathcal{V}_{\alpha\beta}|_{\bi{r}=\mathbf{0}}\!\int_0^{+\infty}\rmd\tau\,\tau^3V(\tau)+\textrm{h.o.t.}\;.
 \end{eqnarray}
 Notice that, for these results to hold, it is crucial to have an always-positive temporal correlation function
 for the turbulent fluctuation. It is known \cite{M97} that different conclusions can be drawn
 if this positivity assumption is loosened, because the presence of negatively-correlated time shifts
 can alter the sign of the temporal integrals under investigation.\\
 As a possible perspective, it would be interesting to extend this analysis from tracers to inertial particles,
 i.e.\ inclusions endowed with non-negligible relative inertia with respect to the carrier fluid,
 typically because of their small-but-finite size and/or different mass density.
 For these particles, an expansion in the Stokes number (if small) can performed along with the multiple-scale procedure
 \cite{PS05,PS07,MAMM12}, yielding an equation for the leading inertia-driven correction to the auxiliary field
 --- and thus to the eddy diffusivity --- very similar to (\ref{ae}) for the operatorial structure on the left-hand side,
 but with a forcing term on the right-hand side extremely more cumbersome and intricated. The extension of the present study to keep
 particle inertia into account is therefore possible from a conceptual point of view, and refrained only because of computational difficulties.

 \ack 
 MMA and SG were partially supported by CMUP (UID/\linebreak[0]MAT/\linebreak[0]00144/\linebreak[0]2013), which is funded by FCT (Portugal)
 with national (MEC) and European structural funds (FEDER), under the partnership agreement PT2020. AM thanks the financial support from the
 PRIN 2012 project no.\ D38C13000610001 funded by the Italian Ministry of Education and from the Italian flagship project RITMARE.

 \appendix
 \section{Calculation details} \label{app}
 
 Let us remind that the Heaviside theta is defined as the primitive of the Dirac delta,
 \begin{equation} \label{ht}
  \theta(y)=\!\int_{-\infty}^y\rmd z\,\delta(z)\qquad(\textrm{with }\theta(0)=\frac{1}{2})\;,
 \end{equation}
 that the definition of the functional derivative is
 \begin{equation} \label{fd}
  \frac{\uD F[f(y)]}{\uD f(z)}=\lim_{\epsilon\to0}\frac{F[f(y)+\epsilon\delta(y-z)]-F[f(y)]}{\epsilon}\;,
 \end{equation}
 and that the Furutsu--Novikov--Donsker theorem \cite{F63,N65,D64} for integration by parts of zero-mean Gaussian fields reads:
 \begin{equation} \label{gi}\fl
  \langle f(\bi{x},t)F[f(\bi{x}',t')]\rangle=\!\int\!\rmd\bi{x}''\int\!\rmd t''\,\langle f(\bi{x},t)f(\bi{x}'',t'')\rangle\left\langle\frac{\uD F[f(\bi{x}',t')]}{\uD f(\bi{x}'',t'')}\right\rangle\;.
 \end{equation}
 For future use, let us compute the first and second functional derivatives of $\bi{w}$ with respect to $\bi{v}$:
 \begin{eqnarray} \label{pd}
  \frac{\uD w_{\beta}(\bi{x},t)}{\uD v_{\sigma}(\bi{x}^*,t^*)}&=&\!\int\!\rmd\tau\,\theta(t-\tau)\bigg\{\!-\delta(\bi{x}-\bi{x}^*)\delta(\tau-t^*)[\partial_{\sigma}w_{\beta}(\bi{x},\tau)+\delta_{\beta\sigma}]\nonumber\\
  &&\left.-[\bi{v}(\bi{x},\tau)\cdot\bpartial+(\bi{U}\cdot\bpartial-\kappa\partial^2)]\frac{\uD w_{\beta}(\bi{x},\tau)}{\uD v_{\sigma}(\bi{x}^*,t^*)}\right\}\nonumber\\
  &=&-\theta(t-t^*)\delta(\bi{x}-\bi{x}^*)[\partial_{\sigma}w_{\beta}(\bi{x},t^*)+\delta_{\beta\sigma}]\\
  &&-\!\int\!\rmd\tau\,\theta(t-\tau)\bi{v}(\bi{x},\tau)\cdot\bpartial\frac{\uD w_{\beta}(\bi{x},\tau)}{\uD v_{\sigma}(\bi{x}^*,t^*)}\nonumber\\
  &&-\!\int\!\rmd\tau\,\theta(t-\tau)(\bi{U}\cdot\bpartial-\kappa\partial^2)\frac{\uD w_{\beta}(\bi{x},\tau)}{\uD v_{\sigma}(\bi{x}^*,t^*)}\nonumber\;,
 \end{eqnarray}
 and
 \begin{eqnarray} \label{sd}\fl
  \frac{\uD^2w_{\beta}(\bi{x},t)}{\uD v_{\sigma}(\bi{x}^*,t^*)\uD v_{\rho}(\bi{x}^{\dag},t^{\dag})}\nonumber\\
  =\!\int\!\rmd\tau\,\theta(t-\tau)\bigg\{\!-\delta(\bi{x}-\bi{x}^*)\delta(\tau-t^*)\partial_{\sigma}\frac{\uD w_{\beta}(\bi{x},\tau)}{\uD v_{\rho}(\bi{x}^{\dag},t^{\dag})}\nonumber\\
  \qquad-\delta(\bi{x}-\bi{x}^{\dag})\delta(\tau-t^{\dag})\partial_{\rho}\frac{\uD w_{\beta}(\bi{x},\tau)}{\uD v_{\sigma}(\bi{x}^*,t^*)}\nonumber\\
  \qquad\left.-[\bi{v}(\bi{x},\tau)\cdot\bpartial+(\bi{U}\cdot\bpartial-\kappa\partial^2)]\frac{\uD^2w_{\beta}(\bi{x},\tau)}{\uD v_{\sigma}(\bi{x}^*,t^*)\uD v_{\rho}(\bi{x}^{\dag},t^{\dag})}\right\}\nonumber\\
  =\theta(t-t^*)\theta(t^*-t^{\dag})\delta(\bi{x}-\bi{x}^*)\times\\
  \qquad\times\{[\partial_{\rho}w_{\beta}(\bi{x},t^{\dag})+\delta_{\beta\rho}]\partial_{\sigma}\delta(\bi{x}-\bi{x}^{\dag})+\delta(\bi{x}-\bi{x}^{\dag})\partial_{\sigma}\partial_{\rho}w_{\beta}(\bi{x},t^{\dag})\}\nonumber\\
  \quad+\theta(t-t^{\dag})\theta(t^{\dag}-t^*)\delta(\bi{x}-\bi{x}^{\dag})\times\nonumber\\
  \qquad\times\{[\partial_{\sigma}w_{\beta}(\bi{x},t^*)+\delta_{\beta\sigma}]\partial_{\rho}\delta(\bi{x}-\bi{x}^*)+\delta(\bi{x}-\bi{x}^*)\partial_{\rho}\partial_{\sigma}w_{\beta}(\bi{x},t^*)\}\nonumber\\
  -\!\int\!\rmd\tau\bigg\{\theta(t-\tau)[\bi{v}(\bi{x},\tau)\cdot\bpartial+(\bi{U}\cdot\bpartial-\kappa\partial^2)]\frac{\uD^2w_{\beta}(\bi{x},\tau)}{\uD v_{\sigma}(\bi{x}^*,t^*)\uD v_{\rho}(\bi{x}^{\dag},t^{\dag})}\nonumber\\
  \quad-\theta(t-t^*)\theta(t^*-\tau)\delta(\bi{x}-\bi{x}^*)\partial_{\sigma}[\bi{v}(\bi{x},\tau)\cdot\bpartial+(\bi{U}\cdot\bpartial-\kappa\partial^2)]\frac{\uD w_{\beta}(\bi{x},\tau)}{\uD v_{\rho}(\bi{x}^{\dag},t^{\dag})}\nonumber\\
  \quad\left.-\theta(t-t^{\dag})\theta(t^{\dag}-\tau)\delta(\bi{x}-\bi{x}^{\dag})\partial_{\rho}[\bi{v}(\bi{x},\tau)\cdot\bpartial+(\bi{U}\cdot\bpartial-\kappa\partial^2)]\frac{\uD w_{\beta}(\bi{x},\tau)}{\uD v_{\sigma}(\bi{x}^*,t^*)}\right\}\nonumber\;.
 \end{eqnarray}

 Equation (\ref{vw}) implies the use of expression (\ref{pd}), which in its turn brings about a recursive substitution.
 Indeed, while the first term on right-hand side of (\ref{pd}) is closed, the third one must be expressed through (\ref{pd}) itself
 --- computed at a different time --- as well as the second one which however is better calculated as
 \begin{eqnarray} \label{ep}\fl
  \!\int\!\rmd\bi{x}^*\!\int\!\rmd t^*\,\langle v_{\alpha}(\bi{x},t)v_{\sigma}(\bi{x}^*,t^*)\rangle\!\left\langle\!\int\!\rmd\tau\,\theta(t-\tau)\bi{v}(\bi{x},\tau)\cdot\bpartial\frac{\uD w_{\beta}(\bi{x},\tau)}{\uD v_{\sigma}(\bi{x}^*,t^*)}\right\rangle\nonumber\\\fl
  =\!\int\!\rmd\bi{x}^*\!\int\!\rmd t^*\int\!\rmd\tau\,\theta(t-\tau)\langle v_{\alpha}(\bi{x},t)v_{\sigma}(\bi{x}^*,t^*)\rangle\!\left\langle v_{\mu}(\bi{x},\tau)\partial_{\mu}\frac{\uD w_{\beta}(\bi{x},\tau)}{\uD v_{\sigma}(\bi{x}^*,t^*)}\right\rangle\nonumber\\\fl
  =\!\int\!\rmd\bi{x}^*\!\int\!\rmd t^*\int\!\rmd\tau\int\!\rmd\bi{x}^{\dag}\!\int\!\rmd t^{\dag}\,\theta(t-\tau)\langle v_{\alpha}(\bi{x},t)v_{\sigma}(\bi{x}^*,t^*)\rangle\times\nonumber\\\fl
  \qquad\times\langle v_{\mu}(\bi{x},\tau)v_{\rho}(\bi{x}^{\dag},t^{\dag})\rangle\partial_{\mu}\!\left\langle\frac{\uD^2w_{\beta}(\bi{x},\tau)}{\uD v_{\sigma}(\bi{x}^*,t^*)\uD v_{\rho}(\bi{x}^{\dag},t^{\dag})}\right\rangle
 \end{eqnarray}
 (i.e.\ by then using (\ref{sd}) in this latter expression). Let us thus analyse how each term of the final result
 (\ref{fr}) is generated, and what is its order of magnitude if the forms (\ref{sc}) are used for the temporal correlation.
 \paragraph*{$\bullet$} The substitution of the first line of (\ref{pd}) gives the first line on the right-hand side of (\ref{fr}).
  This term is $O(T^0)$ for any well-behaved form of $V(\Delta t)$ and is exact for delta-correlated flows,
  while for short-correlated ones it is corrected by the other contributions that we are going to show.
  Notice that here the $T^0$ behaviour is due to the cancellation between the factor $T$ at denominator due to the appearance
  of one instance of the tensor $\langle\bi{v}\bi{v}\rangle$, and the factor $T$ at numerator induced by the presence
  of one temporal integral accompanied by one Heaviside theta.
 \paragraph*{$\bullet$} The second line on the right-hand side of (\ref{fr}) is the leading correction $O(T^1)$ stemming from the
  substitution of the second line of (\ref{pd}). More precisely, it originates from the closed contributions (first four lines)
  on the right-hand side of (\ref{sd}) when replaced in (\ref{ep}), by virtue of the fact that many terms vanish because of
  homogeneity/isotropy/parity (\ref{pi}) or incompressibility (\ref{if}). One of these terms, (\ref{hf}), keeps finite in case the
  assumption of parity invariance is relaxed into (\ref{pb}).
  The substitution of the (unclosed) remaining three lines of the right-hand side of (\ref{sd}) in (\ref{ep}) results in
  contributions which have not been written explicitly in (\ref{fr}) because subleading, as we will prove shortly.
 \paragraph*{$\bullet$} Lines 3$\div$5 on the right-hand side of (\ref{fr}) arise from the substitution of the third line of (\ref{pd})
  into (\ref{vw}). The only surviving closed term after one substitution consists in the third line of (\ref{fr}), and constitutes
  the leading correction at $O(T^1)$ due to molecular diffusivity (dropped in \cite{BCVV95}).\\
  Among the unclosed contributions, let us focus on the one originating from a recursive substitution of the third line of (\ref{pd}).
  After the second replacement, i.e.\ at $O(T^2)$, two closed terms pop up: one is proportional to $\kappa^2$
  but is not interesting because subleading; the other ---
  fourth line of (\ref{fr}) --- represents the leading correction due to the mean flow
  (absent in \cite{BCVV95}), which because of parity only appears at this even order.\\
  A third substitution of the third line of (\ref{pd}), besides a non-relevant subleading term $\propto\kappa^3$,
  allows us to capture the leading correction $O(T^3)$ due to the interplay between mean flow and molecular diffusivity, i.e.\ the
  fifth line of (\ref{fr}).\\
  Notice that these three contributions arise from the coupling with the reflection-symmetric part of the turbulent flow.
  Their couplings with the antisymmetric helical part appear one order later, i.e.\ at $O(T^2)$, $O(T^3)$
  (written explicitly in (\ref{dh}) and (\ref{mh})) and $O(T^4)$ respectively.
 \paragraph*{$\bullet$} To conclude the analysis, we still have to prove three assertions about the subleading character
  of some terms not considered explicitly here. This can be done by means of a simple dimensional analysis.\\
  i) When replacing recursively the second line of (\ref{pd}) into (\ref{vw}), every substitution increases
   by one the order of smallness in $T$. Indeed, to build a quantity with the dimensions of a square length over time
   such as the eddy diffusivity, in this procedure the only available quantity is a combination of factors
   $\langle\bi{v}\bi{v}\rangle$ (always evaluated at fused points, which does not put any spatio-temporal separation into play),
   that has dimensions of square length over square time, along with the appropriate number
   of temporal integrals --- accompanied by $\theta$'s --- or spatial derivatives.
   (Notice that temporal derivatives do not appear because the starting point is the integrated equation (\ref{fi}),
   and that spatial integrals disappear because they are accompanied by the same number of $\delta$'s.)
   With respect to the zeroth order (\ref{dc}), if one introduces further multiplicative factors $\langle\bi{v}\bi{v}\rangle$ for $n$ times,
   one in fact gets a factor $T^{-n}$ from $V$ because of (\ref{sc}), times the $n$-th power of the dimensions of $\mathcal{V}$
   (square length over time). This dimension excess must be compensated not only by $2n$ ``harmless'' spatial derivatives,
   but also by $2n$ temporal integrals, which (as easily realizable e.g.\ for the decaying-exponential form) introduce
   a factor $T^{2n}$. The net result is an increase in the smallness by an order $T^n$. For instance, when comparing the second line
   of the right-hand side of (\ref{fr}) with the first one, the procedure has $n=1$, and indeed there are two more spatial derivatives
   and two more temporal integrals besides one more factor $T$ at denominator, so this contribution is $O(T^{2-1})$ smaller than the previous one.
   Additional recursive substitutions can be discarded, as they would increase $n$.\\
  ii) When replacing recursively the third line of (\ref{pd}) into (\ref{vw}), again every substitution increases
   by unity the order of smallness in $T$. This means that, if we want to explicitly write down only the leading correction
   for each physical effect, we can confine ourselves to: molecular diffusivity alone, i.e.\ third line in (\ref{fr}) at $O(T^1)$;
   mean flow alone, i.e.\ 
   fourth line in (\ref{fr}) at $O(T^2)$; interplay between $\kappa$ and $\bi{U}$, i.e.\ 
   fifth line in (\ref{fr}) at $O(T^3)$. To prove this assertion, it is sufficient to note that now one has two additional
   dimensional quantities available, $\kappa$ (square length over time, which must appear with a positive integer power)
   and $\bi{U}$ (length over time, which must appear with an even power due to parity invariance),
   but that the factor $\langle\bi{v}\bi{v}\rangle$ only appears once, because of the absence of additional applications
   of the Furutsu--Novikov--Donsker theorem besides (\ref{vw}). With respect to the zeroth order given by the first line of (\ref{fr}),
   if $\kappa$ and $\bi{U}$ appear $m$ and $l$ times respectively, a total of $2m+l$ spatial derivatives are implied,
   along with $m+l$ additional temporal integrals. These latter induce a factor $T^{m+l}$ that progressively increases
   the order of smallness, as $V$ appears only once and cannot introduce additional factor $\propto T$ at denominator.\\
  iii) What is left is a mix of cross substitutions of both the second and the third lines of (\ref{pd}) into (\ref{vw}).
   In view of what stated in points i) and ii), they can safely be completely neglected from (\ref{fr}).
   Indeed, all the quantities $\kappa$, $\bi{U}$ and $\langle\bi{v}\bi{v}\rangle$ have time at denominator.
   For every time the molecular diffusivity or the mean flow pop up, they must be accompanied by one temporal integral,
   which increases by unity the order of smallness. For every time the tensorial quantity $\langle\bi{v}\bi{v}\rangle$ pops up,
   it brings about a factor $T^{-1}$ from (\ref{sc}) but also two temporal integrals that increase by two the order of smallness,
   and the net result is again an increase by one.

\end{document}